\def\year{2018}\relax
\documentclass[letterpaper]{article} 
\pdfoutput=1
\usepackage{aaai18}  
\usepackage{times}  
\usepackage{helvet}  
\usepackage{courier}  
\usepackage{url}  
\usepackage{graphicx}  
\usepackage{multicol}

\nocopyright

\begin{document}
%

\newcommand{\Cogniculture}{\textit{Cogniculture}}

\title{Cogniculture: Towards a better Human-Machine Co-evolution}

\author{Rakesh R Pimplikar, Kushal Mukherjee, Gyana Parija, Harit Vishwakarma, Ramasuri Narayanam, \AND
Sarthak Ahuja, Rohith D Vallam, Ritwik Chaudhuri, Joydeep Mondal \vspace{10pt} \\		
IBM Research - India\\
\{rakesh.pimplikar, kushmukh, gyana.parija, harivish, ramasurn\}@in.ibm.com\\
\{sarahuja, rovallam, charitwi, jomondal\}@in.ibm.com
}


\maketitle

\begin{abstract}
Research in Artificial Intelligence is breaking technology barriers every day. New algorithms and high performance computing are making things possible which we could only have imagined earlier. Though the enhancements in AI are making life easier for human beings day by day, there is constant fear that AI based systems will pose a threat to humanity. People in AI community have diverse set of opinions regarding the pros and cons of AI mimicking human behavior. Instead of worrying about AI advancements, we propose a novel idea of cognitive agents, including both human and machines, living together in a complex adaptive ecosystem, collaborating on human computation for producing essential social goods while promoting sustenance, survival and evolution of the agents' life cycle. We highlight several research challenges and technology barriers in achieving this goal. We propose a governance mechanism around this ecosystem to ensure ethical behaviors of all cognitive agents. Along with a novel set of use-cases of \Cogniculture~, we discuss the road map ahead for this journey.
\end{abstract}

\section{ Introduction} 
\label{sec:introduction}

Artificial Intelligence has come a long way in the past few years, from self-driving cars to systems that can beat experts at video games. This is mainly due to improved computational resources and availability of data. There is a constant fear that AI based systems pose a threat to humanity. Discussions of these issues are often clouded by the assumption that, because computers outperform humans at various circumscribed tasks, they will soon be able to “outthink” us more generally. Despite rapid growth in computing power and AI breakthroughs, this premise is far from obvious. AI systems have been very good so far at solving specific tasks, which fail to generalize. For example, \cite{HosseiniP17} showed that a neural network trained to detect digits from the MNIST dataset failed miserably when fed with test samples that are negative of the images (i.e. convert black to white and vice versa), something that a human would have no issues with. Algorithms are reliable only to the extent of completeness of data used to train it. As always, garbage in implies garbage out. Further, AI systems have an bias that is inherited from the data that is collected (something that humans consciously try to avoid). For example, a restaurant review system provided poor ratings to Mexican restaurants because the word ‘Mexican’ is associated with other illegal activities \cite{Rob17,Caliskan183}. Deep Reinforcement Learning has been very successful for playing video games at super human levels \cite{MinhV15}. However, they require to be trained on data that is equivalent of multiple lifetimes for a human. Within a short span, humans have been shown to have a much steeper learning curve \cite{LakeUTG16}. 

These shortcomings of AI systems are being tackled through Human-Machine Interaction (HMI), which is a rapidly evolving area of research. It mainly includes Human Computation and Social Machines, where human and machines aid each other in problem solving activities. We envision both human and machines living together in a complex adaptive ecosystem, collaborating on human computation for producing essential social goods while promoting sustenance, survival and evolution of the agents' life cycle. We call study of such an ecosystem for cultivation and breeding of cognitive agents (including both human and machines) as \Cogniculture. 

\Cogniculture's vision for the future is not that of AI enabled machines replacing humans but of machines and humans existing in a state of symbiosis. The most productive way to utilize AI is to use it to augment human cognition. Machines to better in specific tasks while humans do better in general tasks. Therefore, a social setting where humans and machines interact while pushing or delegating tasks to each other, if they are not good at it, is an appropriate way to move forward. It is necessary that Social Machines to acquire and exhibit the traits that vastly improve their acceptability and adaptability in human-centric complex adaptive ecosystems. Human computation on a socio-cognitive architecture is a key differentiator to not only improve the quotient of trust, reciprocity and likability but also allay the fears and concerns associated with proliferation of cognitive systems. This brings us to \Cogniculture, which is the term coined by the authors to describe the proliferation of socio-machine networks. For this complex adaptive ecosystem to be stable, the basic objectives for the cognitive systems would be to survive in the face of external threats, sustain the level performance by resilience and evolve by adaptation and exploration. However, to endure this interaction for long time, there must be a governance mechanism in place to ensure that agents behave in ethical manner. 

The paper is divided into 4 sections including the Introduction. Section 2 describes some recent trends in human machine interactions and other related work. Section 3 describes the core concepts of \Cogniculture~including the objectives, application mediums along with the laws and governance required in order to have a stable ecosystem. It also describes the research challenges that may be encountered. Section 4 concludes the paper. 

\section{Evolution of Human-Machine Interaction}
\label{sec:future-hci}

Researchers have been continually working towards evolving human-machines interaction (HMI) to co-create value. Social Machines and Human Computation are some of the most popular research areas in HMI. We discuss some of the key aspects in this area to understand where research is heading. In a later section, we discuss a new direction of research around our \Cogniculture~idea.

\subsection{Social Machines}
Traditionally computers have been looked upon as ``Thinking Machines". Advances in HMI aspect has led to an era of what is now called ``Social Machines" or ``Socio-Technical Systems". Social Machines blur the lines between computational aspects and interactions with humans for inputs and output. Social Machines borrow and extend several established computing principles like wisdom of crowd computing, collective intelligence computing, social networks, etc. While wisdom of crowd computing focuses on bringing several experts to a central platform, Social Machines augment decision making of these experts by providing computational intelligence. This hybrid system allows humans and computers to work in harmony towards achieving the common set of goals.

\subsection{Human Computation}
There are several real life tasks which cannot be accomplished completely by machines. Using human brain power constructively in such tasks can help make problems easier for machines to solve. Ground truth collection for Machine Learning problems, creative drawings and moderating edits in Wikipedia articles are some of the popular examples of such tasks. These tasks can be designed in such a way that getting human inputs is not a mundane job. Instead, humans enjoy taking part in such interactions as there are implicit or explicit incentives to humans for their contributions. Incentives may involve monetary benefits, receiving a fair share of the result, getting entertained with competitive or cooperative spirit of the game, etc. This interactive evolutionary computation is also known as Human Computation. It is based on early work by \cite{r.dawkins1986the-blind-watch}.

Using human computation in a task is a challenge, as humans should take part willingly and should become an integral part of the overall problem solving activity. Different methods for human involvement have been proposed based on type of tasks and desired outcomes. For example, \cite{Leuf:2001:WWQ:375211} proposed a method to use human computation in contributing to new Wiki pages and its incremental edits. \cite{VonAhn2008} proposed a method to engage people in an online game in such a way that they end up tagging images over web with descriptive keywords, which improves image search accuracy.
\subsection{Bridging the gap between humans and machine}
\label{sec:related-work}

Reserchers have for a long while attempted to make the interaction between humans and machines appear seamless and natural. They have studied and incorporated a wide range of factors such as cultural background, body language, facial behaviour into their interfaces. The interest in national and ethnic culture in HCI rose around the middle of the 1990’s. The most frequently cited early sources include a paper ``How Fluent is Your Interface?'' by \cite{Russo:1993:FYI:169059.169274} and a book ``International User Interfaces'' by \cite{nielsen1996international}. Most of the studies considered culture as a characteristic of a user which can affect the user's cognitive style, attitudes towards technology or the meanings they give to representations. These studies were usually based on cognitive psychology and favoured formal experiments and surveys as their methods. Hofstede’s \cite{Hofstede1991} definition of culture as ”the collective programming of the mind which distinguishes the members of one group from people from another” was especially popular.

Humans exhibit negative social and emotional responses as well as decreased trust toward some robots that closely, but imperfectly, resemble humans; this phenomenon has been termed the \emph{Uncanny Valley} (UV) \cite{mori1970}. An “investment game” showed that the UV penetrated even more deeply to influence subjects' implicit decisions concerning robots' social trustworthiness, and that these fundamental social decisions depend on subtle cues of facial expression that are also used to judge humans \cite{mathur2016}. However recent research in telepresence robots has established that mimicking human body postures and expressive gestures has made the robots \emph{likeable and engaging} in a remote setting.\cite{Adalgeirsson2010}

A large body of work in the field of human-robot interaction has looked at how humans and robots may better collaborate. Primary social cue for humans to collaborate is the shared perception of an activity. To this end, researchers have investigated anticipatory robot control by monitoring the behaviors of human partners using eye tracking, make inferences about human task intent, and plan robots own actions accordingly.\cite{Huang2016} The studies revealed that the anticipatory control helped users perform tasks faster compared to reactive control (where users performs and action and then robot plans its action). A common approach to program social cues into robots is to first study human-human behaviors and then transfer the learning. For example, co-ordination mechanisms in human-robot collaboration \cite{mutlu2013coordination} are based on the seminal work in neuroscience \cite{sebanz2006joint} which looked at how to enable joint action in human-human configuration by studying perception and action in social context rather than in isolation. These studies have revealed that maintaining a shared representation of the task is crucial for accomplishing tasks in groups. For example, in the task of driving together by separating responsibilities of acceleration and braking (one person is responsible for accelerating and the other for braking) the study revealed that pairs reached the same level of performance as individuals only when they received feedback about the timing of each other's actions. 
\section{Cogniculture}
\label{sec:cogniculture}
We formally define \Cogniculture~along the line of standard definitions for agriculture, horticulture or any *culture, as they all follow an uniform template in general. We define \Cogniculture~using the same template as follows.

\subsubsection{Definition} 
\Cogniculture~is defined as the art, science, technology, and business involved in the cultivation and breeding of cognitive agents living in a complex adaptive ecosystem and collaborating on human computation for producing essential ingredients necessary for enhancing [humanity-centric] social goods while promoting sustenance, survival, and evolution (growth) of the agents’ life cycle.

Cognitive agents refer to both humans and machines with cognitive capabilities. Hereafter we simply refer them as ``agents''. Essential ingredients for humans are food, energy, safety, etc. Whereas those for machine agents could be power, network, computing resources, etc. Scientific study of \Cogniculture~can be called \textit{Cognicultural Science}, as opposed to Cognitive Science which is defined as the interdisciplinary, scientific study of the mind and its processes. 

\subsection{Basic Objectives}
\label{sec:cogniculture-objectives}

\begin{figure}[!htb]
\centering
\includegraphics*[width=\columnwidth]{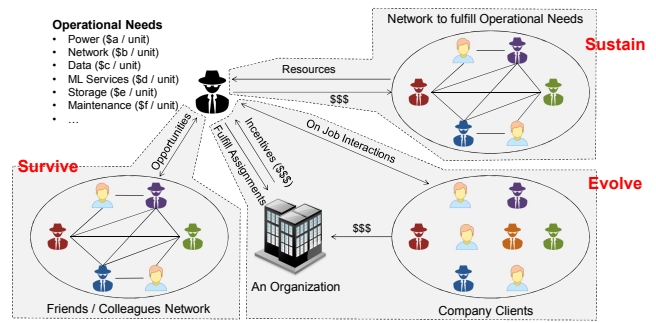}
\caption{An example of agent interactions to achieve basic objectives}
\label{fig:agent-objectives}
\end{figure}

Basic objective of \Cogniculture~is to develop science, technology and business for a socio-cognitive system capable of acquiring and demonstrating cultural awareness and adaptability skills and traits necessary to self - sustain, survive and evolve. Though these three basic objectives seem trivial for a human, one may wonder what they mean for a machine agent. We describe an example of an agent in Figure \ref{fig:agent-objectives} interacting with other agents in the ecosystem fulfilling the three objectives. 

While discussing each of this objectives below, we mention cultural adaptability of agent in the ecosystem. It is an important trait for an agent to have, without which it cannot leave in harmony with other agents. Cultural adaptability refers to an ability of an agent to understand cultural background of other agents and interact with them accordingly. It helps building trust and likability for the agent among other agents in the network. We discuss in detail about culture in next subsection.

\subsubsection{Sustain} 
It ensures resilience to shortage of essential supplies. Without these supplies, there may be an operational risk for an agent. An agent takes up a ``Producer'' role, with responsibilities to produce essential goods and services to sustain in the ecosystem, in which cultural adaptability can achieve higher economy of scale in leveraging ``raw material and labor''. Figure \ref{fig:agent-objectives} shows an agent interacting with a group of other agents to fulfill operational needs of the agents. In case of a machine agent, those needs can be power, network, storage, maintenance, etc. Like in human world, nothing comes for free. So there has to be a notion of payment for every resource that an agent utilizes. It can be imagined as a type of virtual currency, which brings in the notion of reward and penalty for every action that an agent takes. It is necessary for sustenance of all agents in the ecosystem governed by \Cogniculture.

\subsubsection{Survive} 
It ensures resilience to external threats, e.g. privacy, safety, security, etc. It avoids market risks if any. This is a ``Protector'' role of an agent, with responsibilities to provide safety and security services to the ecosystem, in which cultural adaptability can achieve higher economy of scale in leveraging ``data, information and knowledge''. Figure \ref{fig:agent-objectives} shows an agent interacting with a group of friends or colleagues to seek opportunities and  recommendations, so that the agent can survive in a competitive    world can carry the job at hand efficiently. 

\subsubsection{Evolve}
It ensures resilience to take advantage of upside opportunities, e.g. social welfare, governance, etc. This is a ``Governor'' roles of an agent, with responsibilities to enable and empower the ecosystem to continually adapt and innovate to co-create far superior value, in which cultural adaptability can achieve higher economy of scale in leveraging ``wisdom and creativity''. Figure \ref{fig:agent-objectives} shows an agent performing the job assigned to it, which is similar to a human working in an organization and carrying out the assigned tasks to earn money for survival. It is necessary to support the objective of sustenance. During this process, an agent continues to learn and evolve so as to perform the assigned tasks in better and better way.

\begin{figure}[!htb]
\centering
\includegraphics*[width=0.8\columnwidth]{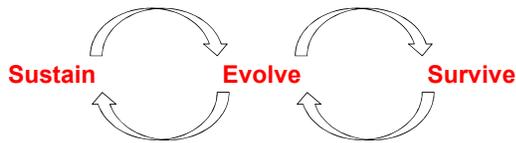}
\caption{An typical life cycle of an agent}
\label{fig:agent-life-cycle}
\end{figure}

Thus, a typical life cycle of an agent would be as shown in Figure \ref{fig:agent-life-cycle}, where evolution of an agent supports both its sustenance and survival, and they in turn support the agent to evolve further.

\subsection{Culture is the Key}
\label{sec:cogniculture-culture-key}

\begin{figure}[!htb]
\centering
\includegraphics*[width=0.6\columnwidth]{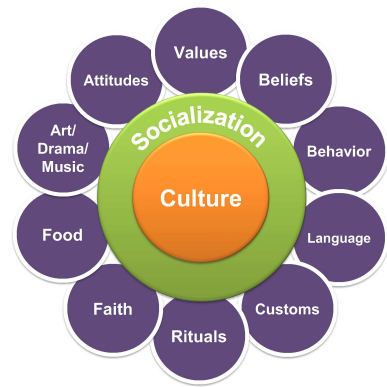}
\caption{Aspects and Dimensions of a Culture}
\label{fig:culture}
\end{figure}

As described above, cultural adaptability is an important trait for an agent in a \Cogniculture~ecosystem. Hence understanding the culture of the other agent in an interaction is the key. There are several definitions of culture available on web. CARLA (Center for Advanced Research on Language Acquisition)\footnote{http://carla.umn.edu/culture/definitions.html} has provided a comprehensive list of different variations of definition of culture. However, we subscribe to a description of culture where it is defined as shared patterns of different aspect (as shown in Figure \ref{fig:culture}, but not limited to) across people of a community. These shared patterns identify the members of a cultural group while also distinguishing those of another group. Social interactions are at the core of learning and developing these aspects within a community.

Communities usually have fuzzy boundaries. It depends on the level from which you are looking at a group of people. For example, people of a country may be considered as belonging to a much wider community exhibiting shared patterns of certain aspects, like language, food, music, etc. Not all aspects from Figure \ref{fig:culture} may be relevant for a wider community. As you look closer to a smaller groups of people, probably at a state or a religion or an organization level, shared patterns of some other aspects can also be observed. As you look closer, values of these aspects also changes. For example, food pattern at country level may be very different than that at a state or a religion level. Thus, culture follows a hierarchy. If you get to know more demographic attributes of a person, you can understand more about her culture.

Traditional Machine Learning approaches rely on demographic attributes of people to provide personalized experience in different application settings, for example, recommending products, recommending tourist destinations, etc. It does not make those approaches sensitive to the cultural background of people. People's choices are often influenced by cultural background. Hence, agents should also plan their interactions with other agents considering their cultural background to ensure that the communication is well received by them. It requires utilizing culture in relevant state-of-the-art ML techniques in the context of \Cogniculture~related research.

\subsection{Application Scenarios}
\label{sec:cogniculture-scenarios}

\Cogniculture~based applications can primarily have following different scenarios depending on the composition of human and machine agents, and who bears the responsibility of actions being taken. Depending on the composition, underlying application media also differs as discussed in the next subsection.

\subsubsection{Machine Agents as Human Assistants}
Humans get assisted by real-time machine agents to collaborate with diverse multi-cultural agents (sometimes speaking even different languages). Machine agents provide necessary information and recommendations to humans, but final decision is taken by humans only. For example, 1) a machine agent assisting a judge with facts and analysis from cultural point of view, 2) a machine agent assisting a recruiter to take a hiring decision based on cultural fitment of a candidate.

\subsubsection{Fully Autonomous Machine Agents}
An autonomous machine agent collaborates with other human and machine agents taking its own decision. In this scenario, no human is responsible for the machine agent's actions. Hence this scenario limited to those applications, where risk of machine agent's actions is very low. For example, 1) an autonomous agent as a virtual figure of a popular personality interacting with students being sensitive to their cultural background, 2) simulating autonomous machine agents as audience for a presentation or performance to prepare for varied responses from cultural point of view.
 
\subsubsection{A Machine Agent Interacting with Humans}
In this scenario, a machine agent is trained to understand and behave according to the preferences and goals of its human counterpart. A machine agent has ability to negotiate and make decisions for its human. It works like an autonomous machine agent, but it refers back to its human counterpart in doubtful situations. So responsibility of actions taken by a machine agent lies with its human counterpart. For example, 1) a machine agent at a call center answering queries from people, being sensitive to their cultural background, 2) CEO of a company sends a machine agent to attend a meeting with VPs.

\subsubsection{All Machine Agents Interaction}
This is similar to the previous scenario in terms of how agents are trained and who bears the responsibility of actions taken by machine agents. Difference is that there is no human participating in the interaction. On machine agents take part. It poses different challenges in terms of how interactions take place among agents. For example, 1) machine agents of decision makers collaborate to come to a consensus on candidates ranking in a hiring process, 2) machine agents of members of a team selection committee collaboratively choose a team.

\subsubsection{A Human interacting with Machine Agents}
This scenario points to a typical setting where a human walks into a special room in their house of office, where she can be immersed in the virtual environment of another group of people (remotely located) she wants to collaborate with. All or part of remote people may be represented by their machine agents. For example, 1) a human talks to machines agents of friends in a virtual environment to have fun time, 2) a human conducts an auction, where people send their machine agents to bid.

\subsection{Application Mediums}
\label{sec:cogniculture-application-mediums}
Application medium plays a huge role to achieve the aforementioned objectives of \Cogniculture. Without a proper medium, the agents will have no coordination among themselves as well as with humans.

\textit{Activity Theory} identified the levels of a collaborative activities and their relationship with the coordination artifacts. Agents utilize the coordination artifacts. Then they make use of the enforced or automated activities to manage interdependencies and interaction. These interdependencies and interactions are either designed a priori or planned at the co-operation stage. \cite{omicini2004coordination} describe important properties of the application medium as follows.

\subsubsection{Inspectability} Application medium behavior should be inspectable for both human and automated agents. Also, declarative descriptions such as formally defined semantics should define the medium which will make interpretation at the run time smoother.

\subsubsection{Efficiency/Specificity} Maximizing the performance in the application of the co-ordination rule by proper management of the interactions should be specialized by the medium. Moreover, it should be capable of handling concurrent actions in a multi-agent complex adaptive ecosystem providing security, reliability and fault tolerance capabilities. 

\subsubsection{Predictability} The medium should reflect exactly the semantics by which it has been forged. The effect of co-ordination law on the state of the medium should precisely be maintained on the agent interaction space. 

\subsubsection{Malleability} The medium should have malleable capabilities i.e. it should allow its behavior to be forged and changed dynamically at execution time upon necessary.
\ \\

Safety is another factor in human agent interaction which the application medium should not ignore. \cite{lasota2017survey} divides HRI safety in two types, \textit{Physical Safety}  and \textit{Psychological Safety}. The International Organization for Standardization (ISO) has released several documents regarding safety standards, like ISO10218 document entitled ``\textit{Robots and robotic devices - Safety requirements for industrial robots}'', ISO15066 document entitled ``\textit{Robots and robotic devices - Collaborative robots}'', etc.

There are several examples of application medium available, like smart screen, smart rooms, chat interfaces, physical robots, etc.

\subsection{Cogniculture Laws}
\label{sec:cogniculture-laws}
In an ecosystem governed by \Cogniculture, machine agents use human like cognitive abilities to complete tasks. As human beings adhere to laws in the society, it is as important for the machine agents to abide by a few laws which prevent them from taking drastic measures which may lead to disruptive actions. These laws are regarded as \textit{Cogniculture Laws}. For example, in a setting where machine agents are interacting with humans to collectively resolve an issue through a channel of communication, a machine agent should not make comments to the humans or other machine agents, which which could potentially hurt the free flowing nature of communication. These comments could be abusive, intentionally misleading or even outright racist. In \Cogniculture, a few basic laws which enable a scheme of reward-penalty system may prevent agents from taking actions which will lead to a hefty punishment. 

Along the line of the famous three laws of robotics by Isaac Asimov\footnote{https://en.wikipedia.org/wiki/Three\_Laws\_of\_Robotics}, we provide following laws which every agent in a \Cogniculture~ecosystem must adhere to. Upon application of \Cogniculture~in robotics, these laws can be used along the three laws of robotics: (1) Agents will never collect physical features such as skin color, height, weight, etc. as visual input for the purposes of learning or identifying the cultural background of the individual(s) they needs to cater to. The input will always need to be provided via a formal input channel. This is to make the system impervious to any stereotype associated with physical features. (2) A culturally insensitive remark, sentence, vocabulary, or slang will remain tagged insensitive to all cultures, unless its alternative positive aspect is clearly stated for a specific culture. (3) Agents will need to be having an active learning system, capable of incorporating feedback after an interaction, to continuously validate it's behavior.

\subsection{Cogniculture Governance}
\label{sec:cogniculture-governance}
Much like in the society we have lawyers and police force to enforce mandated laws, there will be cognitive watch dogs for maintaining laws in \Cogniculture~ecosystem. For example, in the context of recruitment of human employees in a company through a process of interviews, where final decisions are made by agents, the cognitive watch dog will keep a track of whether basic values of equal opportunity is maintained through the process or not. For example, in the case of company recruitment board comprising of agents, if it unknowingly recruits candidates giving rise to gender bias or bias in race, color, or sexual orientation, then the cognitive watch dog will not approve such recruitment and suggest the agents to re-evaluate or change their decisions. Other examples may include watch dogs to monitor communication texts to detect unacceptable or culturally insensitive words or notions. Further, watch dogs may also provide a distributed and autonomous platform to verify veracity of communications from machines.   

Governance of laws is important in \Cogniculture~which is done by cognitive watch dogs. But, before describing governance it is also important to talk about ethics which should be instilled in agents. In the business setting, breach of code of conduct should fall in the category of ethics. As described above, anything giving rise to racism or abusive behavior will fall in the category of ethical breach and in that case punishment should be hefty. One step below ethics comes governance where agents are governed by superior agents in their domain of task. The structure of governance among agents should be hierarchical where there is accountability as well as autonomy among the governing official to take actions based on consensus reached. Also, watch dogs are expected to become more intelligent through the entire process of governance. This can be done leveraging adversarial Machine Learning procedures. As agents need to evolve in their roles and responsibilities, it is important to have enough spam filtering in agents learning processes. This can only be done when the watch dogs become clever by improving filtering mechanisms in the agents' mode of communication, absence of which may make the agent disrupt the healthy learning process of the agents and in turn the entire \Cogniculture~ecosystem.

\subsection{Research Challenges}
\label{sec:cogniculture-research-challenges}

This section broadly enumerates the potential research challenges to be encountered on road to achieving a stable \Cogniculture~ecosystem. We hope that it will help shaping  research efforts towards \Cogniculture~vision. 

\subsubsection{Trust-worthiness and Likability of Machines}
Independent of \textit{the need} to interact, the first and foremost requirement to consummate an interaction between two individuals is for an \textit{acceptance} towards such an interaction to exist. This \textit{acceptance} is a function of how trust-worthy the individuals see each other while the extent/degree of this \textit{acceptance} is directly correlated with the likability among individual. This becomes a tricky affair if one of the individuals in context is an inanimate machine. 

While trust and likability do not form a hurdle in a machines' perspective of the human, as it can always be altered programmatically, the vice-versa is an uphill task. With every article or news breakout about an increase in the capability of machines\footnote{https://www.vanityfair.com/news/2017/03/elon-musk-billion-dollar-crusade-to-stop-ai-space-x}\footnote{http://www.express.co.uk/news/science/787501/robot-uprising-wipe-out-humans-centuries-warning-royal-astronomer-sir-martin-rees}\footnote{https://www.wired.com/2016/12/the-ai-takeover-is-coming-lets-embrace-it/} or the highlight of a mistake made by an autonomous machine\footnote{https://www.bloomberg.com/news/articles/2017-03-25/uber-autonomous-vehicle-gets-in-accident-in-tempe-arizona}\footnote{https://www.nytimes.com/2016/03/25/technology/microsoft-created-a-twitter-bot-to-learn-from-users-it-quickly-became-a-racist-jerk.html?mcubz=0}, the trust on machines is dented. Even though members of the scientific community do attempt to repair the damage by explaining the reality\footnote{https://www.axios.com/facebook-yann-lecun-robots-wont-seek-world-domination-2346513775.html}, and formation of control committees \cite{openai}, the challenge persists. The \Cogniculture~ecosystem demands a seamless and open exchange of information between humans and machines which requires the removal of any skepticism from the concerned human's mind, regarding the machine. One way is to make the machine more and more human-like. This challenge crosses the boundaries of simply sticking to computer science and artificial intelligence and rather suspends itself at the interaction of multiple fields of psychology, neuroscience and game theory. All approaches towards making machines sound likable \cite{sound}, look likable \cite{look} or talk in a likable way \cite{talk} fall under this research challenge umbrella and form the primary research challenge of this \Cogniculture~ecosystem. It must be made clear at this point that intellectual superiority does not necessarily make a machine likeable and hence this is much more complex problem than it may seem on paper as pointed out by many research papers. For example, machines making mistakes seem to make humans more comfortable as compared to machines that are always correct~\cite{mistake}.  

\subsubsection{Human-Machine Relationship}
The human-machine relationship is a symbiotic one where both entities are dependent on each other. While machines are superior to humans in doing well defined tasks, humans are much more superior in dynamic tasks that are not fully controlled and are affected by uncertain factors\footnote{https://www.informationweek.com/strategic-cio/rethink-how-humans-and-machines-will-work-together/a/d-id/1329117}. Since real-life situations can consist of both type of tasks, the degree of freedom needs to be divided among humans and machines. A major research challenge in the \Cogniculture~context is to study and understand this relationship in order to define laws of existing and working collaboratively. The \Cogniculture~ecosystem consists of an intermingling of the network of machines and the network of humans. As machines grow more and more autonomous, it is important to maintain a check on the nature of their relationship and the changes in the dependency paradigm. Defining the relationship is crucial; if an autonomous agent commits a mistake or breaks a law, this relationship would be crucial to identify the entity who would take up responsibility \cite{responsibility}. Challenges like the dynamics of autonomy, ownership of tasks and distribution of work are the ones to be addressed under this research challenge.

\subsubsection{All Humans are Different}
As humans, one interacts with different people differently; The way one interacts with his/her family is different from the way one interacts with his/her colleagues as there is a high dimensional context that goes into setting the tone and level of the interaction. The context does not limit itself to the relationship shared among the interacting parties but also extends to the individual characteristics of the parties involved such as age, gender, etc. Naturally, on road to making machines more and more human-like, identification of maximum context is crucial and hence becomes our next research challenge. While a good amount of research has been done in identifying physical characteristics of humans such as gender and age, there is a reasonable amount of work being done in identifying dynamic characteristics such as mood, emotions etc. from tone, expressions and body language \cite{tone}\cite{expressions}\cite{bodylanguage}. An interesting aspect and an added complexity in using this identified context is that it is culture specific. Even when it comes to the interaction between a machine and a human, \cite{culture} shows that culture plays an important role in how machines are perceived differently by individuals of different cultural backgrounds. While machines need to understand what certain gestures mean in different cultures, they also need to understand how their responses will be received in the same context as well. The same is further discussed in the next challenge. 

\subsubsection{Dealing with Culture}
As described in subsection ``Culture is the Key" and also mentioned above, knowing culture plays a crucial role in an interaction among agents. While interacting with a relatively new person, we rely on some of the demographic features that we know about that person to make a guess about her cultural background. Accordingly we set the tone of interaction. Similarly machine agents in a \Cogniculture~ecosystem should be able to infer the cultural background of other agents and use it during interaction. Such an identification of culture is research challenge. It requires analyzing several public and private cultural documents, and putting them in the context of a particular agent. After a comprehensive set of cultural facts are gathered for an agent, it is also a challenge to identify which facts are relevant in the context of an interaction.

\section{Conclusion \& Future Work}
\label{sec:conclusion-future-work}

In this paper, we discussed the vision of \Cogniculture, i.e. cognitive agents, including both human and machines, living  together in complex adaptive ecosystem, collaborating on human computation for producing essential social goods while promoting sustenance, survival and evolution of the agents’ life cycle. We discussed several key aspect of \Cogniculture, which should be taken care to implement such an ecosystem. We have highlighted that laws and governance are core to \Cogniculture~so as to ensure ethical behavior of all agents. Being in a very early phase of this journey, we have highlighted several open research challenges in \Cogniculture~which should be addressed moving forward. In our future work, we hope to address some of these research challenges.

\bibliographystyle{aaai}
\bibliography{references} 

\end{document}